\documentclass[10pt]{iopart}
\usepackage[driverfallback=dvipdfm,bookmarks=true,colorlinks,citecolor=blue,linkcolor=blue,anchorcolor=blue,filecolor=blue,urlcolor=blue,]{hyperref}
\usepackage{graphicx}
\usepackage{subfigure}
\begin{document}

\title[Jena et al.]{Exploring the impact of $\Delta$--isobars on Neutron Star}

\author{ Rashmita Jena,$^{1,2}$, S.K. Biswal,$^2$, Padmalaya Dash,$^{3}$
R.N. Panda,$^{3}$
and M. Bhuyan, $^{4}$}
\address{$^{1}$Department of Physics, Fakir Mohan University, Balasore, Odisha, 756019, India\\
$^{2}$Department of Physics, K. K. S. Women's College, Balasore, Odisha, 756001, India\\
$^{3}$Department of Physics, SOA University, Bhubaneswar, Odisha, 751003, India\\
$^{4}$Department of Physics, Institute of Physics,  Bhubaneswar, Odisha, 751005, India}

\ead{jrashmitaphy@gmail.com}
\ead{subratphy@gmail.com}
\vspace{10pt}

\begin{abstract}

We include the $\Delta$--isobars in the equation of state (EOS) of neutron star (NS) and study its effects with various parameter sets of the RMF model. We compare our results with the NS's constraints from the mass--radius measurement of PSR J0348+0432, PSR J1614-2230, PSR J0030+0451, PSR J0740+6620, PSR J0952-0607, and tidal deformability of GW170817. We calculate the mass--radius profile and tidal deformabilities of the NS using 21 parameter sets of the RMF model.Analyzing the result with various parameters, it is clear that only few parameter sets can satisfy simultaneously the constraints from NICER and GW170817. NLD parameter set satisfy all the constraints of NICER and GW170817. For its strong predictive power for the bulk properties of the neutron star, we take NLD parameter set as a representative for the detailed calculation of effect of $\Delta$--isobar on neutron star properties. We demonstrate that it is possible that $\Delta$--isobar can produce at 2--3 times the saturation density by adjusting the coupling constants $X_{\sigma\Delta}$, $X_{\rho\Delta}$ and $X_{\omega\Delta}$ in an appropriate range. Bulk properties of the NS like mass--radius profile and tidal deformability is strongly affected by the interaction strength of $\Delta$--isobar.
Our calculation shows that it is also possible that by choosing $X_{\sigma\Delta}$, $X_{\rho\Delta}$ and $X_{\omega\Delta}$ to a suitable range the threshold density of $\Delta^-$--isobar become lower than $\Lambda^0$ hyperon. For a particular value of $\Delta$--coupling constants, the $R_{1.4}$ decrease by 1.7 km. This manuscipt give an argumentative justification for allowing $\Delta$--isobar degrees of freedom in the calculation of the NS properties. 
\end{abstract}

\vspace{2pc}

\ioptwocol

\section{Introduction}
Neutron stars are compact objects
 characterized by their high density, small radius,
 and strong gravitational pull compared to ordinary stars, first introduced by Walter Baade and Fritz Zwicky in 1934 \cite{Baade34,shapiro83,glenb97}. These
 celestial bodies emerge from the remnants of
 massive stars following type II supernova explosions. In 1939, Oppenheimer and
 Volkoff conducted the inaugural calculation on neutron
 star structures, hypothesizing that their cores consist of densely packed neutrons
 resembling ideal gases.
 Initially, due to their diminutive size and the limitations of optical telescopes in detecting their thermal radiation
 across vast distances, neutron stars did not attract much scientific interest. This changed in 1967 when Jocelyn Bell Burnell and Anthony Hewish
 discovered the first radio pulsar, PSR B1919+21 \cite{bell68}, notable for its pulse period of 1.337s and a pulse width of 0.04s,
 sparking renewed attention and extensive research to study the composition and structure of neutron stars.

Primarily, it is believed that a neutron star (NS) 
 is made up of  90\% neutrons and
 10\% protons. However, further research has 
 revealed that there are the possibilities of hyperon formation
 \cite{amba60,glen85,glen91}, kaon condensation \cite{glen85,neha12,neha13,glen98,glen99,chri00}, and deconfinement of quark matter
 \cite{coll75,xia2018,dexh18} in the core of the NS. Significant research has explored
 hyperon creation at densities 2--3 times of nuclear matter density within 
the NS core. Yet, the impact of the $\Delta$--isobars on the NS structure
 has received
 comparatively less attention due to its association with high-density regions.
 However, recent research has shown that by adjusting the coupling
 constant, the $\Delta$--isobars
 can appear at a density of 2--3 times that of nuclear saturation 
 density, creating
 a $\Delta$--isobars puzzle similar to the hyperon puzzle 
 \cite{torsten10,Drago14,Sahoo18,spinella18,LI18,Ribes19,Malfatti19,Dex21,Rather2023}.
Within $\beta$--stable matter, the $\Delta$--isobars have the potential to 
emerge first by substituting a neutron and an electron at the Fermi
Sea's peak. The presence of the $\Delta$--isobars
softens the equation of state (EOS), consequently reducing the maximum mass and
radius of the NS \cite{torsten10,Drago14,Sahoo18,LI18,Ribes19,Dex21, Marczenko22, Rather2023}. It is reported that the $\Delta$--isobar can reduce the canonical radius $R_{1.4}$ \cite{torsten10,LI18}. In ref. \cite{LI18}, the authors shows, the attractive potential of $\Delta$--isobars reduce the $R_{1.4}$ up to 2 km. Presence of $\Delta$ also lowers the canonical
tidal deformability $\Lambda_{1.4}$ \cite{Li19, Marczenko_2022}. In ref. \cite{Marczenko_2022}, the authors have found the canonical radius of pure nucleonic state decreases from 1.4 km to approximately 2 km for $m_0^N$ = 700 MeV and 600 MeV respectively, when $m_0^N = m_0^{\Delta}$. They also found the value of $\Lambda_{1.4}$ reduces for smaller value of $m_0^{\Delta}$ due to the earlier onset of $\Delta$--isobar. For $m_0^N = m_0^{\Delta}$ the $\Lambda_{1.4}$ reduce significantly.
Furthermore, in the ref. \cite{Marquez22}, the authors investigate the properties of deltic stars and neutron stars admixed of nucleons, hyperons, and $\Delta$--isobars. They demonstrate that the dominance of w--mesons contributes to a stiffer equation of state (EOS), which results in larger maximum mass and higher speed of sound in nucleon, hyperon, and delta admixed neutron stars at the core centre.  In neutron stars with this admixture, the presence of $\Delta$--isobars is limited to less than 20\% at the centre. At densities between 2 to 3 times the nuclear saturation density, their presence rises to below 30\%. Conversely, in deltic stars, the proportion of $\Delta$--isobars at the core centre can reach up to 80\%.

In this study, we delve into the effects of the $\Delta$--isobars on 
various properties of the neutron stars (NS), focusing on 
their maximum mass, canonical radius, and tidal deformability.
We incorporate the $\Delta$--isobar into the equation of state (EOS)
and use a range of parameter sets of the relativistic mean-field
(RMF) model, namely HS \cite{horo81}, LA \cite{furn87}, H1 \cite{hadd99}, LZ \cite{rein89}, NL3 \cite{lala97}, NL3* \cite{lala09},
NL3--II \cite{lala97},
NL1 \cite{rein86}, NL--RA1 \cite{rash01}, NLD \cite{furn87}, NL--SH \cite{SHARMA93}, IFSU \cite{fatt10}, IFSU* \cite{fatt10}
SINPB \cite{mond16},
GM1 \cite{glen91}, GL97 \cite{glenb97}, GL85 \cite{glen85}, G3 \cite{kumar18}, IOPB \cite{kumar18}. 
We aim to establish an EOS that satisfies the NICER (Neutron Star Interior Composition Explorer) constraints mainly PSR J0348+0432: M = 1.97 $M_\odot$ to 2.05 $M_\odot$ \cite{Antoniadis13}, PSR J1614-2230: M = 1.892 $M_\odot$ to 1.924 $M_\odot$ \cite{Arzoumanian18}, PSR J0030+0451: M = 1.18 $M_\odot$ to 1.49 $M_\odot$, R = 11.52 km to 
13.85 km \cite{Riley19}, PSR J0740+6620: M = 2.006 $M_\odot$  to 2.139 $M_\odot$, R =  11.41 km to 13.69 km \cite{Riley21},
and the mass of the heaviest NS: PSR J0952-0607: M = 2.18 $M_\odot$ to 2.52 $M_\odot$ \cite{Romani22}, and the GW170817 constraint. The dimensionless canonical tidal deformabilty  is $\Lambda_{1.4}\le 800$ as mentioned in \cite{abbo17}, however it is again reported that the range of canonical tidal deformability should be $\Lambda_{1.4}$=70--580 for the EOS which lie within 90\% credible level for the GW170817 event  \cite{abbo18}.

However, none of the above parameter sets can fully align with all the constraints imposed by NICER and the GW170817 observations. We select the NLD parameter set
 as it is most compatible with the NICER and GW170817 constraints. For detail analysis of $\Delta$--isobar on neutron star's bulk properties we use NLD parameter set and vary the $\Delta$--coupling constants  $X_{\sigma_\Delta}$, $X_{\omega\Delta}$ and $X_{\rho_\Delta}$ within a particular range. We show that the variations in the coupling constants $X_{\sigma_\Delta}$, $X_{\omega\Delta}$ and $X_{\rho_\Delta}$ affect the threshold density of hyperons. By adjusting the $\Delta$--coupling constants, it is possible to emerge the $\Delta$--isobars mainly $\Delta^-$ at lower density in comparison to $\Lambda^0$. Also, the presence of $\Delta$--isobars affect the EOS, mass, canonical radius ($R_{1.4}$), and canonical tidal deformability ($\Lambda_{1.4}$) of the neutron star. 

This manuscript is organised as follows: started with a brief introduction to the effect and importance of $\Delta$--isobar in hyperon  
star calculation. In theoretical formalism part, we discuss about the theoretical frame work for the calculation of mass, radius, and tidal 
deformability. A section is devoted to discuss the results of our calculation. Finally, the manuscript is ended 
with a summary and concluding remarks. 

\section{Theoretical Framework}
\subsection{Relativistic Mean Field Theory}
The relativistic mean field (RMF) model provides more convenient way to
explore the internal structure of the NS. In 
RMF theory, nucleons are 
interact via exchange of mesons and 
photons. These interacting mesons are categorized on the basis of their spin, isospin and parity. Generally, the scalar or vector meson and isoscalar or isovector meson with natural parity are chosen. More specifically, the isoscalar--scalar meson $\sigma$
(J = 0, T = 0), the isoscalar--vector mesons $\omega$ (J = 1, T = 0), and the isovector--vector mesons $\rho$ (J = 1, T = 1) are considered. The isoscalar-scalar meson $\sigma$ 
 facilitates medium--range 
attraction among nucleons, while the isoscalar--vector mesons $\omega$ are 
responsible for short-range repulsion. 
The isovector--vector mesons $\rho$ characterize isovector properties. The photon 
governs electromagnetic interaction. In this calculation, the interaction of the
strange scalar meson $\sigma^*$ and the 
strange vector meson $\phi$ with 
hyperons are taken care similarly to the $\sigma$ meson and 
$\omega$ meson interaction with nucleons \cite{cava08,bipa89,bedn03,spal99,bunt04,scha96,feng92}.

The effective Lagrangian for 
RMF model is \cite{rein86,mill72,furn87,Bed2000,ring96,Biswal19}\\
\begin{eqnarray}
&&{\cal L}=\sum_B\overline{\psi}_B\bigg(
i\gamma^{\mu}\partial_{\mu}-m_B+g_{\sigma B}\sigma -g_{\omega B}\gamma_\mu
 \omega^ \mu \nonumber \\
&&-\frac{1}{2}g_{\rho B}\gamma_\mu\tau\rho^\mu-g_{\phi_0 B}  \gamma_{\mu} {\phi_0}^{\mu} \bigg)
\psi_B + \frac{1}{2}\partial_{\mu}\sigma\partial_{\mu}\sigma \nonumber \\
&&-m_{\sigma}^2\sigma^2
\left(\frac{1}{2}+\frac{\kappa_3}{3!}\frac{g_{\sigma}\sigma}{m_B}
+\frac{\kappa_4}{4!}\frac{g_{\sigma}^2\sigma^2}{m_B^2}\right)
 - \frac{1}{4}\omega_{\mu\nu}\omega^{\mu\nu} \nonumber \\
&&+\frac{1}{2}m_{\omega}^2
\omega_{\mu}\omega^{\mu}\left(1+\eta_1\frac{g_{\sigma}\sigma}{m_B}
+\frac{\eta_2}{2}\frac{g_{\sigma}^2\sigma^2}{m_B^2}\right)
-\frac{1}{4}R_{\mu\nu}R^{\mu\nu} \nonumber \\
&&+\frac{1}{2}m_{\rho}^2
R_{\mu}R^{\mu}\left(1+\eta_{\rho}
\frac{g_{\sigma}\sigma}{m_B} \right)
+\frac{1}{4!}\zeta_0 \left(g_{\omega}\omega_{\mu}\omega^{\mu}\right)^2 \nonumber\\
&&+\frac{1}{2} {m_\phi}^2 {\phi_\mu}{\phi^\mu}+\sum_l\overline{\psi}_l\left(
i\gamma^{\mu}\partial_{\mu}-m_l\right)\psi_l \nonumber\\
&&+\Lambda_v R_{\mu}R^{\mu}
(\omega_{\mu}\omega^{\mu}),
\end{eqnarray}
where B use for baryons decuplet n, p, $\Lambda$, $\Sigma^-$, 
$\Sigma^0$, $\Sigma$, $\Xi^-$, $\Xi^0$ and $\Delta$-isobar 
($\Delta^-, \Delta^0, \Delta^+, \Delta^{++}$) 
, $\omega_{\mu\nu}$ and $R_{\mu\nu}$ represent  field tensors
for the $\omega$ and $\rho$ meson fields respectively and are defined as 
$\omega_{\mu\nu}=\partial_\mu \omega_\nu-\partial_\nu \omega_\mu$ and
$R_{\mu\nu}=\partial_\mu R_\nu-\partial_\nu R_\mu$.
\begin{figure}
	\includegraphics[width=1.1\columnwidth, height=8cm]{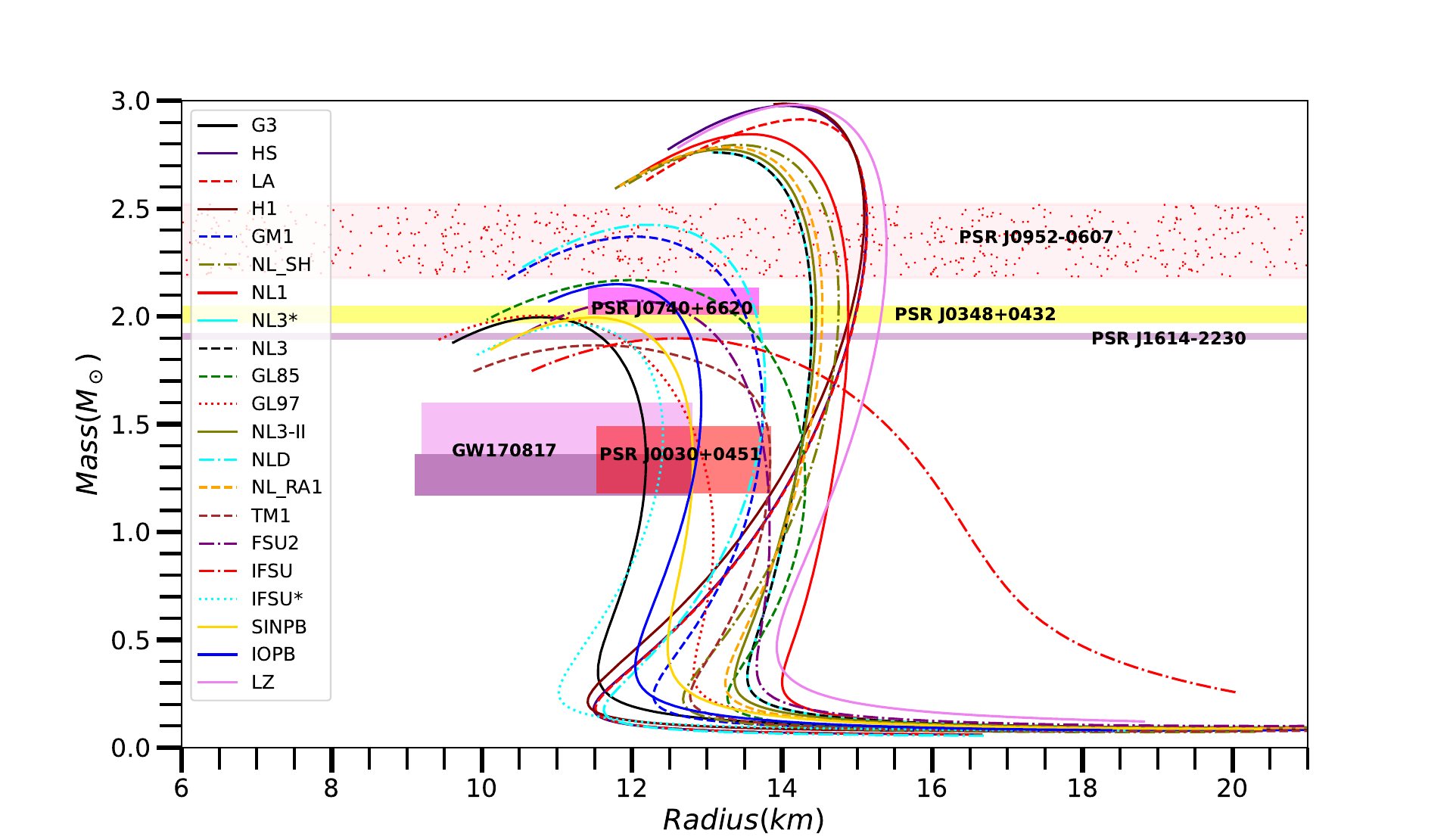}
   \caption{Mass-radius profile of the neutron star with various
	parameter sets of the RMF model. The boxes represent the constraints of NICER and GW170817.}
    \label{fig1}
\end{figure}

By using classical Euler-Lagrangian equation
we get the following equation of motion for the different mesons
\begin{flushleft}
\begin{eqnarray}
m_{\sigma}^2 \left(1+\frac{g_{\sigma N}\kappa_3\sigma_0}{2m_B}
+\frac{\kappa_4 g_{\sigma N}^2\sigma_0^2}{6m_B^2} \right) \sigma_0
-\frac{1}{2}m_{\rho}^2\eta_{\rho}\frac{g_{\sigma N}\rho_{03}^2}{m_B}\nonumber \\
-\frac{1}{2}m_{\omega}^2\left(\eta_1\frac{g_{\sigma N}}{m_B}
+\eta_2\frac{g_{\sigma N}^2\sigma_0}{m_B^2}\right)\omega_0^2
=\sum g_{\sigma B}\rho^s_B.\nonumber\\
\end{eqnarray}
\end{flushleft}
\begin{flushleft}
\begin{eqnarray}
m_{\omega}^2\left(1+\eta_1\frac{g_{\sigma N} \sigma_0}{m_B}
+\frac{\eta_2}{2}\frac{g_{\sigma N}^2\sigma_0^2}{m_B^2}\right)\omega_0 
+\frac{1}{6}\zeta_0g_{\omega N}^2\omega_0^3\nonumber\\
=\sum g_{\omega B}\rho_B.\nonumber\\
\end{eqnarray}
\end{flushleft}
\begin{eqnarray}
 m_{\rho}^2\left(1+\eta_{\rho}\frac{g_{\sigma N}\sigma_0}{m_B}\right)R_0
=\frac{1}{2}\sum g_{\rho B}\rho_{B3}.
\end{eqnarray}
\begin{eqnarray}
{m_\phi}^2 \phi_0 =\sum g_{\phi B} \rho_B.
\end{eqnarray}

To form inside the core of the neutron star, the baryons and leptons must satisfy the $\beta$-stable equilibrium and charge neutrality conditions.
In star matter the nucleons, hyperons, $\Delta$--isobars and leptons satisfy the following $\beta$--equilibrium conditions.
\begin{eqnarray}
    \mu_p = \mu_{\Sigma^+} = \mu_n - \mu_e ,
 \cr   \mu_n = \mu_{\Sigma^0} = \mu_{\Xi^0} ,
 \cr   \mu_{\Sigma^-} = \mu_{\Xi^-} = \mu_n + \mu_e,
  \cr  \mu_{\mu} = \mu_e,
  \cr  \mu_{\Delta^-} = 2\mu_n - \mu_p,
  \cr   \mu_{\Delta^0} = \mu_n, 
  \cr   \mu_{\Delta^+} = \mu_p, 
  \cr   \mu_{\Delta^{++}} = \mu_p - \mu_n. \nonumber
\end{eqnarray}

The charge neutrality condition is satisfied as
\begin{eqnarray}  
n_p + n_{\Sigma^+} + 2n_{\Delta^{++}} + n_{\Delta^+} = n_e + n_{\mu^-} + n_{\Sigma^-} + n_{\Xi^-} + n_{\Delta^-} \nonumber
\end{eqnarray}

Total
energy $\cal E$ density and pressure $\cal P$ can be calculated from 
 energy--momentum tensor $T^{\mu\nu}$ defined as
 \begin{eqnarray}
 T^{\mu\nu} = \sum_{\alpha}\frac{\partial\cal L}{\partial 
 (\partial_{\mu}\phi_{\alpha})}\partial^{\nu}\phi_{\alpha}-\eta^{\mu\nu}\cal L.       
 \end{eqnarray}
\begin{figure}
	\includegraphics[width=1.1\columnwidth, height=8cm]{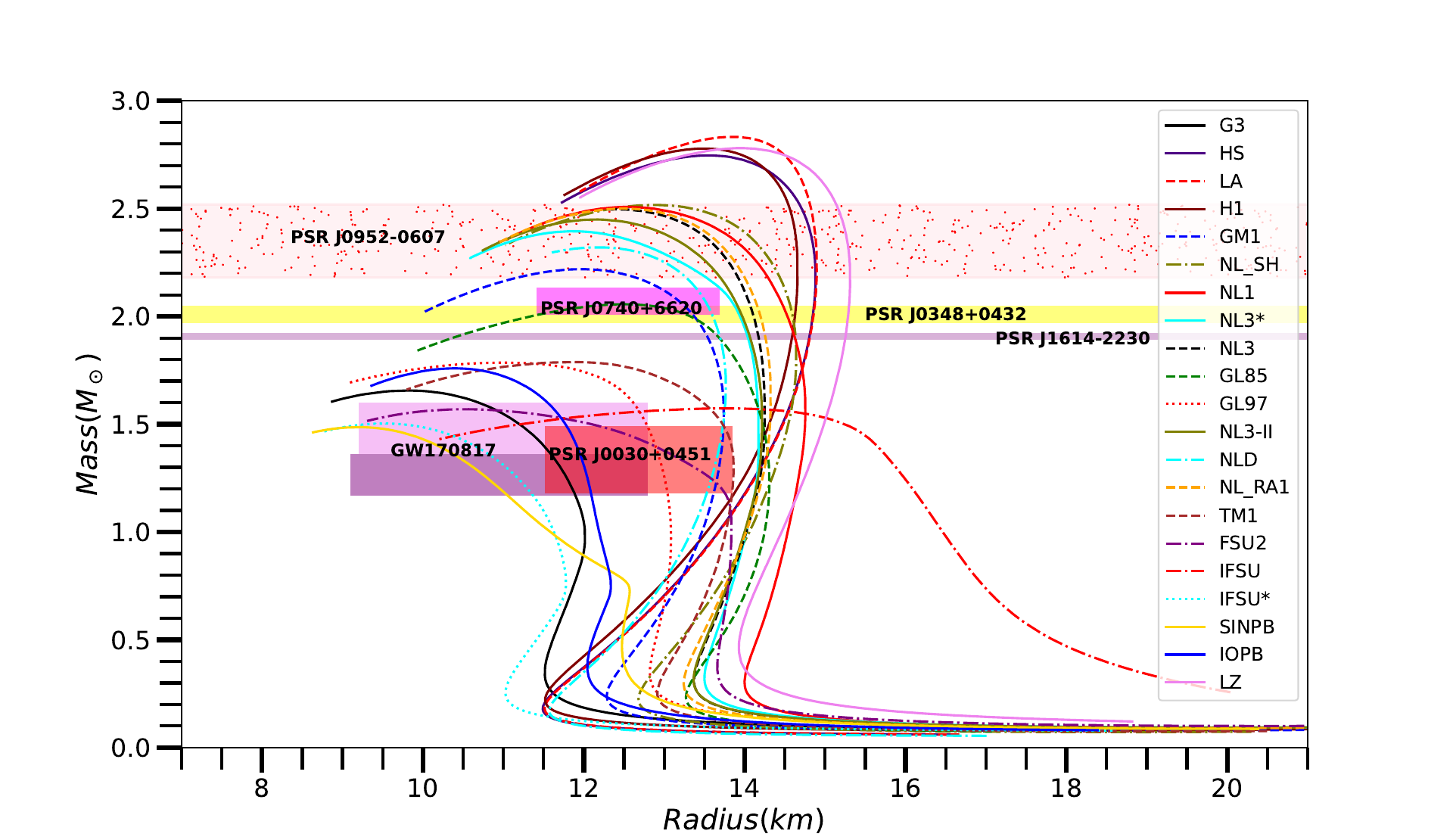}
   \caption{Effect of the $\Delta$--isobar on the mass-radius profile of the
neutron star with the coupling constants $X_{\sigma\Delta}$ = 1.1, $X_{\omega\Delta}$ = $X_{\rho\Delta}$ = 1. The boxes represent the constraints of NICER and GW170817.}
    \label{fig2}
\end{figure}

 The calculated expression for energy density and pressure are given by
\begin{eqnarray}\label{energy}
&&{\cal E}=\sum_B\frac{Y_B}{(2\pi)^3}\int_0^{k_F^B} d^3k \sqrt{k ^2+{m_B}^*}
+\frac{1}{8}\zeta_0g_{\omega}^2\omega_0^4 \nonumber \\
&& + m_{\sigma}^2\sigma_0^2\left(\frac{1}{2}+\frac{\kappa_3}{3!}
\frac{g_{\sigma}\sigma_0}{m_B}+\frac{\kappa_4}{4!}
\frac{g_{\sigma}^2\sigma_0^2}{m_B^2}\right) \nonumber \\
&& + \frac{1}{2}m_{\omega}^2 \omega_0^2\left(1+\eta_1
\frac{g_{\sigma}\sigma_0}{m_B}+\frac{\eta_2}{2}
\frac{g_{\sigma}^2\sigma_0^2}{m_B^2}\right) \nonumber \\
&& + \frac{1}{2}m_{\rho}^2 \rho_{03}^2\left(1+\eta_{\rho}
\frac{g_{\sigma}\sigma_0}{m_B} \right)+\frac{1}{2}{m_\phi}^2 \phi^2
 \nonumber \\
&&+\sum_l \int_0^{k_F^l} \sqrt{k^2+{m_l}^2} k^2 dk
+3\Lambda_V \omega_0^2 R_0^2,
\end{eqnarray}
and
\begin{eqnarray}\label{pressure}
&&{\cal P}=\sum_B\frac{Y_B}{3(2\pi)^3}\int_0^{k_F^B}\frac{k^2 d^3k}{\sqrt{k^2+{m_B^*}^2}}
+\frac{1}{24}\zeta_0g_{\omega}^2\omega_0^4 \nonumber \\
&& - m_{\sigma}^2\sigma_0^2\left(\frac{1}{2}+\frac{\kappa_3}{3!}
\frac{g_{\sigma}\sigma_0}{m_B}+\frac{\kappa_4}{4!}
\frac{g_{\sigma}^2\sigma_0^2}{m_B^2}\right) \nonumber \\
&& + \frac{1}{2}m_{\omega}^2 \omega_0^2\left(1+\eta_1
\frac{g_{\sigma}\sigma_0}{m_B}+\frac{\eta_2}{2}
\frac{g_{\sigma}^2\sigma_0^2}{m_B^2}\right) \nonumber \\
&& + \frac{1}{2}m_{\rho}^2 \rho_{03}^2\left(1+\eta_{\rho}
\frac{g_{\sigma}\sigma_0}{m_B} \right)+\Lambda_V R_0^2 \omega_0^2 \nonumber \\
&&+\frac{1}{3\pi^2}\sum_l \int_0^{k_F^l} \frac{k^4 dk}{\sqrt{k^2+m_l^2}},
\end{eqnarray}
where $l$ stands for the leptons (electrons and muons).
The equation of state describes the variations of the total energy density and 
pressure of nuclear matters with baryon density. These equation of states are used in the TOV equation to calculate bulk properties of the neutron star. 
\begin{figure}
	\includegraphics[width=1.1\columnwidth, height=8cm]{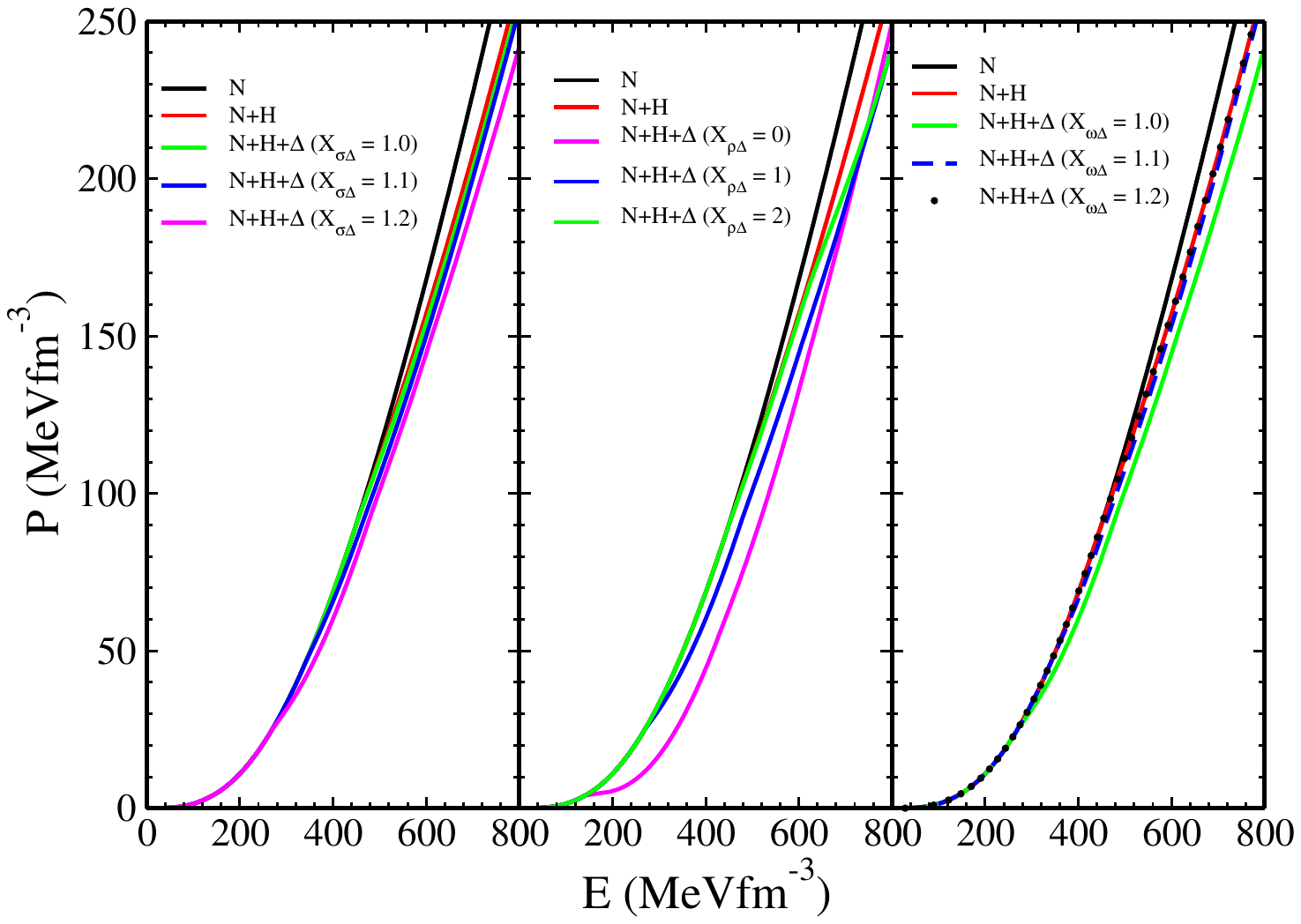}
   \caption{Left panel of the figure shows variation of pressure with energy density for different values of $X_{\sigma\Delta}$ with $X_{\omega\Delta}$ = $X_{\rho\Delta}$ = 1, the middle panel of the figure shows variation of pressure with energy density for different values of $X_{\rho\Delta}$ with $X_{\sigma\Delta}$ = 1.1, $X_{\omega\Delta}$ = 1  and right panel of the figure shows variation of pressure with energy density for different values of $X_{\omega\Delta}$ with $X_{\sigma\Delta}$ = 1.2, $X_{\rho\Delta}$ = 1.} \label{fig3}
\end{figure}
\subsection{Tolman-Oppenheimer-Volkoff  Equation}
Tolman-Oppenheimer-Volkoff (TOV) 
equation 
is used to delineate
the characteristics of neutron stars. 
Specifically, it is deployed to describe the cold, spherically symmetric, and non-
rotating astrophysical bodies like neutron star. Within 
the context of this work, we use the TOV equation to calculate and analyze the 
mass and radius of a neutron star. The TOV equations are given as \cite{tolm39,oppe39,glenb97} 

\begin{eqnarray}
\frac{\partial P}{\partial r}= - \frac{(P+\epsilon(r))(m(r)+4\pi r^3 P)}{r(r-2m(r))},
\end{eqnarray}

\begin{eqnarray}
\frac{\partial m}{\partial r}=4\pi r^2 \epsilon(r),
\end{eqnarray}

\begin{eqnarray}
m(r) =  \int_{0}^{r} 4 \pi r^2 \epsilon(r) \,dr, 
\end{eqnarray} 
where m(r) represent the mass enclosed within the radius r. P(r) and $\cal E$ represent the pressure and total energy density. 
This endeavor is crucial for unraveling the complex structure and dynamics of neutron stars, thus providing invaluable insights into their physical properties and the extreme conditions present within their confines.

\subsection{Tidal Deformability}
Tidal deformability is a property of a neutron star that measure the degree of deformation caused by the gravitational pull of its companion star. This property is affected by the internal structure and equation of state of the NS. 
When a neutron star and its companion star orbit each other, they experience tidal forces due to their orbital motion, which become stronger as they approach to a merger. During this event, gravitational waves are emitted. The first-ever gravitational wave signal from a neutron star--neutron star merger, known as GW170817, was detected by the LIGO and VIRGO collaboration on August 17, 2017 \cite{abbo17,abbo18}. This binary neutron star merger event provides a constraint on the canonical neutron star's tidal deformability within the range of 70--580. The tidal deformability $\lambda$ of a neutron star depends on its radius and tidal Love number $k_2$ \cite{Hinderer08}, which are related by the equation \cite{hind10,kumar17,tuhin18,Raithel18}
$\lambda =\frac{2k_2 R^5}{3}$
and the dimension less tidal deformability is
$\Lambda = \frac{2k_2 }{3C^5}$,
where C = M/R is the compactness of NS and R is the radius of NS. The detail calculation of the tidal deformability can be found in \cite{hind10,kumar17}
\begin{figure}
	\includegraphics[width=1.1\columnwidth, height=8cm]{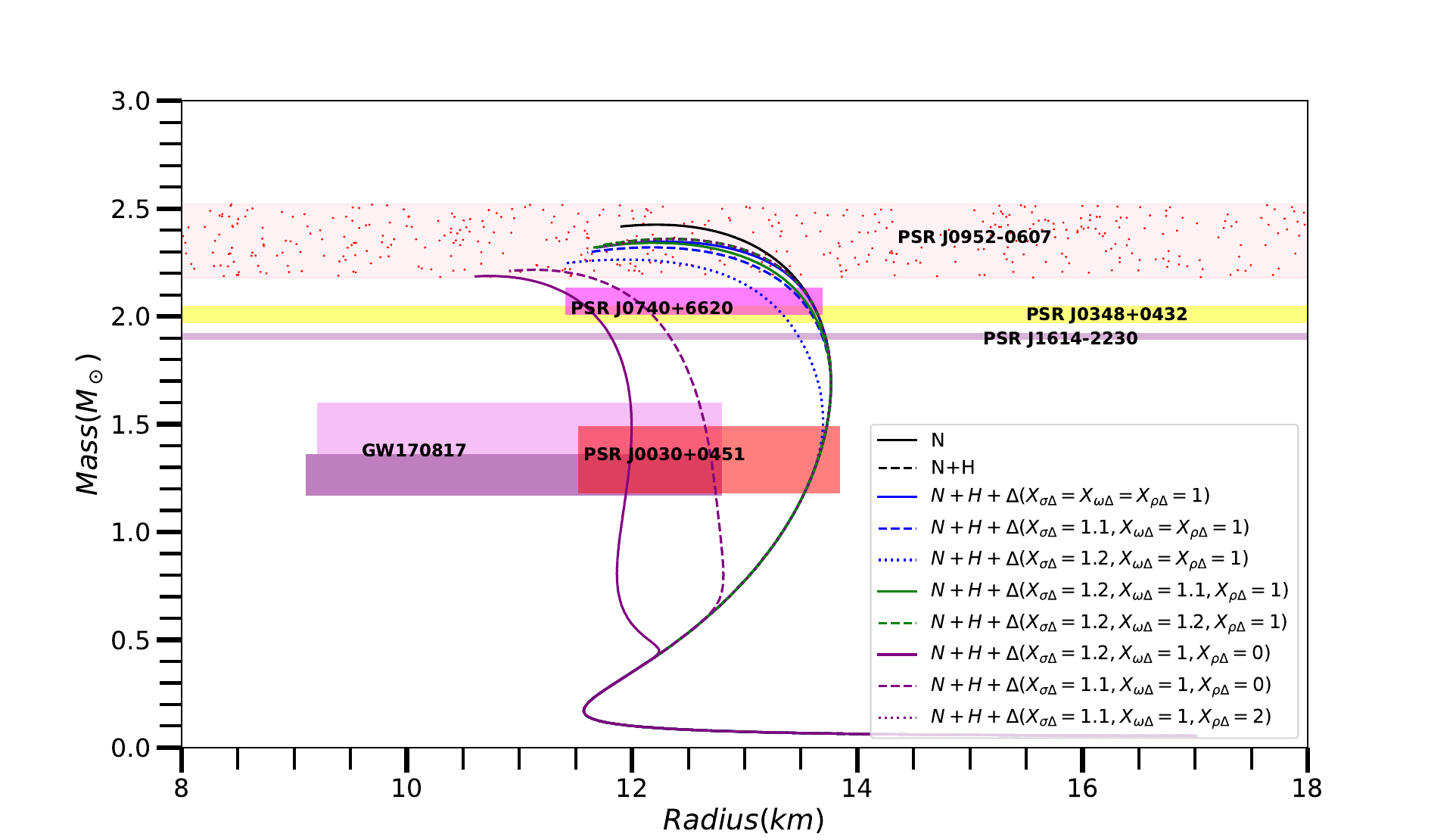}
   \caption{Effect of the $\Delta$--isobar on the mass-radius profile of the
NLD parameter set with the coupling constants $X_{\sigma\Delta}$ = 1--1.2, $X_{\omega\Delta}$ = 1--1.2 and $X_{\rho\Delta}$ = 0--2. The boxes represent the constraints of NICER and GW170817.}
    \label{mr}
\end{figure}

\section{Result and Discussion}
We use 21 different parameter sets of the RMF model to investigate which parameter sets align with NICER and GW170817 constraints. Fig.\ref{fig1} displays the mass-radius plot of these parameter sets with NICER and GW170817 constraints. Analyzing Fig.\ref{fig1} we note that, among all the parameter sets none of them can completely satisfy the NICER and GW170817 constraints. However, GM1 and NLD parameter sets satisfy all the constraints of the NICER and the G3, GL97 and IFSU* parameter sets best align with GW170817 constraints.

Further, we include the $\Delta$--isobars in our EOS to study the effect of the $\Delta$--isobars on neutron star's properties. Initially, it is believed that the $\Delta$--isobars generally appear at higher density. However, further research reveals the possibilities of the $\Delta$--isobars appearance at lower densities. The threshold density of the $\Delta$--isobars can be shifted to a lower density by adjusting their coupling constants in certain range. We follow the procedure to set the coupling constants of different hyperons and $\Delta$--isobars by fixing their potentials to a particular value.

We set the coupling constant to generate the potential  
${U_{\Lambda}} = -28$ MeV \cite{batt97,mill88,tor17,lope18},
${U_{\Sigma}} = +30$ MeV \cite{kohn06,hara05,hara06,frie07,scha00},
 and ${U_{\Xi}}= -18 $ MeV \cite{glen91,scha00} by using the equation
\begin{eqnarray}
U_{Y} = {m_n}(\frac{m_n^*}{m_n}-1)X_{\sigma Y}+(\frac{g_\omega}
{m_\omega})^2\rho_0 X_{\omega Y},
\end{eqnarray}

where Y stands for the different hyperons 
($\Lambda, \Sigma, \Xi $) and $\Delta$--isobars ($\Delta^-, \Delta^{0}, \Delta^+, \Delta^{++}$). 
$X_{\sigma Y}$ and $X_{\omega Y}$ are the coupling constants of the
hyperon--meson and $\Delta$--meson interactions, and $\rho_0$ is the saturation density.
The hyperon--meson coupling constants $X_{\sigma Y}$ and $X_{\omega Y}$ 
are fitted to produce potential depth for various hyperons following the 
assumption that the hyperon--meson coupling constants must be less than the 
nucleon--meson coupling constants. We use  SU(6) symmetry for setting interaction strength of
hyperon and $\rho$ meson interaction \cite{dove84,scha94}:
$X_{\Lambda\rho}=0$, $X_{\Sigma\rho}=2$, $X_{\Xi\rho}=1$.
The interation strength of hyperon and $\phi$--meson are set to be as:
$X_{\phi\Lambda}=- \frac{\sqrt{2}}{3}g_{\omega N}$,  $X_{\phi\Sigma}=
-\frac{{\sqrt{2}}}{{3}}$$g_{\omega N}$, $X_{\phi\Xi}=-\frac{2\sqrt{2}}{3}$$g_{\omega N}$.
We chose the $\Delta$--isobars coupling constants by ensuring that the potential ${U_{\Delta}}$ lies in the range -150 MeV$\leq U_{\Delta} \leq$ -50 MeV and the coupling constants $X_{\sigma_\Delta}$ and $X_{\rho_\Delta}$ satisfy the condition $0\leq X_{\sigma\Delta} - X_{\rho_\Delta}\leq 0.2$ \cite{WEHRB89,Drago14}.
Using this condition, we set $X_{\sigma\Delta} = 1.1$, $X_{\omega\Delta} = 1$ and $X_{\rho\Delta} = 1$ for various parameter sets to calculate the mass and radius as shown in Fig.\ref{fig2}.
\begin{figure}
	\includegraphics[width=1.1\columnwidth, height=8cm]{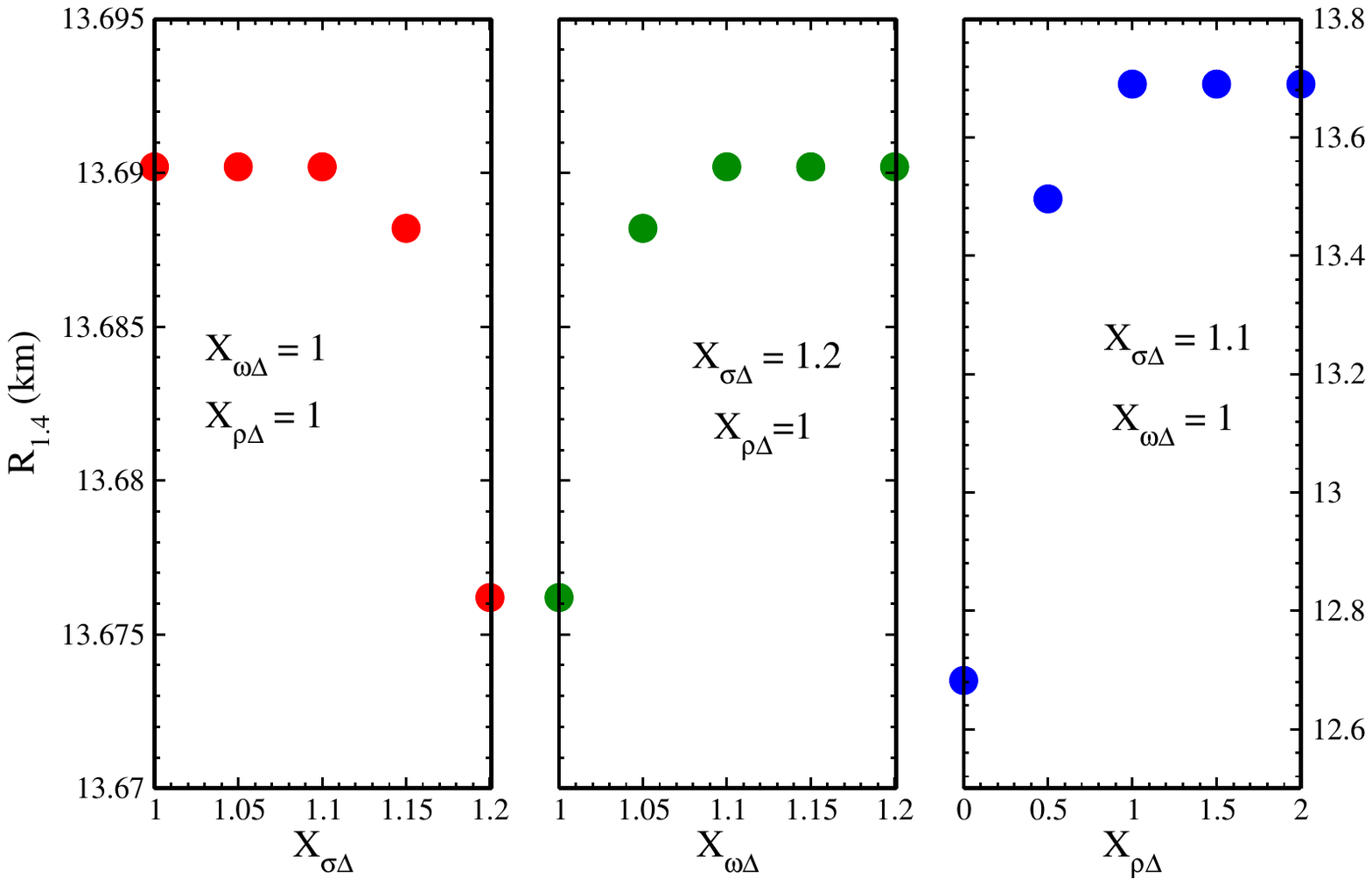}
   \caption{Variation of canonical radius R$_{1.4}$ with X$_{\sigma\Delta}$, X$_{\rho\Delta}$ and X$_{\omega\Delta}$}
\label{fig4}
\end{figure}

Fig.\ref{fig2} shows, the maximum mass and radius for each parameter set decreases due to the presence of $\Delta$--isobar. However, the parameter sets NLD and GM1 are still satisfy the NICER constraints (see Fig.\ref{fig2}). Furthermore, along with G3, GL97, and IFSU* parameter sets, the parameter sets FSU2, IOPB and SINPB also satisfy the GW170817 constraints.
Along with the study of mass and radius, we also want to study how the EOS, canonical radius, and tidal deformability are affected with different values of 
$\Delta$--isobars coupling constants. However, it is difficult to study these properties using all the mentioned parameter sets of the RMF model. Therefore, we choose only NLD parameter set for further studies as it satisfies all the NICER constraints.

For detail calculation, we set the coupling constants of $\Delta$--isobar to different values and study their effect on the EOS of hyperon star as shown in Fig.\ref{fig3} .
In the left panel of Fig.\ref{fig3} we set $X_{\omega\Delta}$ = $X_{\rho\Delta}$ = 1 and vary $X_{\sigma\Delta}$ from 1 to 1.2. The increasing value of $X_{\sigma\Delta}$ makes the EOS softer in higher density due to the attractive potential of $\sigma$--meson. In the middle panel of Fig.\ref{fig3} we fix $X_{\sigma\Delta}$ and $X_{\omega\Delta}$ as 1.1 and 1 respectively and vary $X_{\rho\Delta}$ from 0 to 2. We observe the EOS becomes stiffer for the higher values of $X_{\rho\Delta}$and this effect almost starts from lower density. The stiffness of EOS for higher values of $X_{\rho\Delta}$ is due to the increase in repulsive interaction of $\Delta$--isobars near saturation density. In the right panel of Fig.\ref{fig3}, we set the coupling constants $X_{\sigma\Delta}$ and $X_{\rho\Delta}$ to 1.2 and 1, respectively and vary $X_{\omega\Delta}$ = 1--1.2. The EOS becomes stiffer at higher densities as the $X_{\omega\Delta}$ increases. The increase in $X_{\omega\Delta}$ increases the repulsive interaction of $\Delta$--isobars, leads to the stiffer EOS.
\begin{figure}
	\includegraphics[width=1.2\columnwidth, height = 10cm]{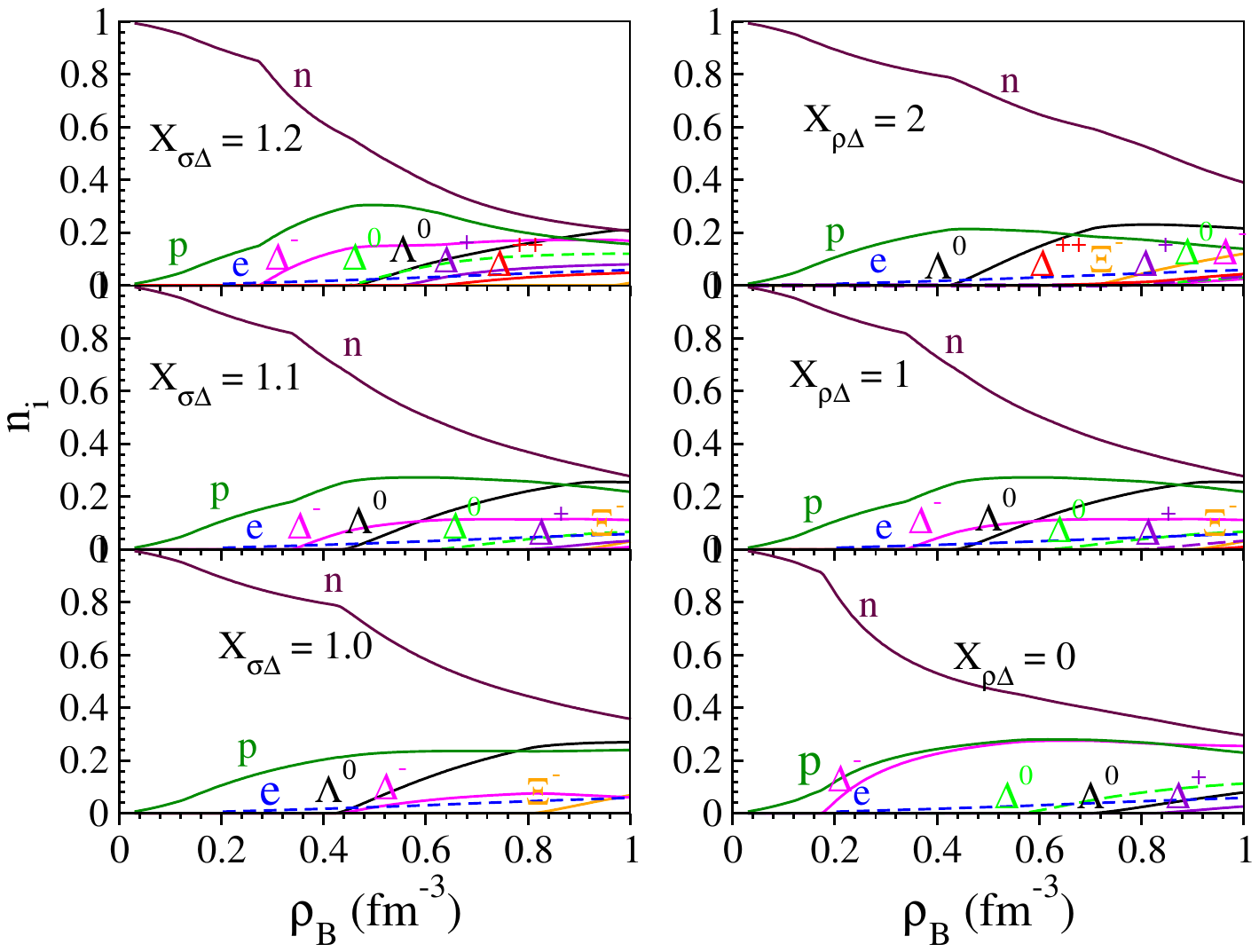}
   \caption{Left panel shows the baryon formation with different values of X$_{\sigma\Delta}$ with X$_{\omega\Delta}$ = 1 and X$_{\rho\Delta}$ = 1 of NLD parameter set. Right panel shows baryon formation of NLD parameter set for different value of X$_{\rho\Delta}$ with X$_{\sigma\Delta}$ = 1.1, X$_{\omega\Delta}$ = 1. Here $\rho_B$ represent the baryon density and $n_i$ repersent the particle fraction.} 
\label{fig5}
\end{figure}

The mass--radius profile of the NLD parameter set with varying $\Delta$--coupling constants is illustrated in Fig.\ref{mr}. We observe that the maximum mass of the neutron star decreases when hyperon degrees of freedom are added to the equation of state (EOS). Furthermore, when incorporating $\Delta$--isobars into the EOS with the coupling constants set to $X_{\sigma\Delta}$ = $X_{\omega\Delta}$ = $X_{\rho\Delta}$ = 1, the maximum mass is reduced. To explore the effect of $\Delta$-isobars, we keep $X_{\omega\Delta}$ and $X_{\rho\Delta}$ fixed at 1 and vary $X_{\sigma\Delta}$ from 1 to 1.2. In this scenario, increasing $X_{\sigma\Delta}$ from 1 to 1.2 leads to a reduction in the maximum mass due to the softening of the EOS. Similarly, by fixing $X_{\sigma\Delta}$ at 1.2 and $X_{\rho\Delta}$ at 1 while varying $X_{\omega\Delta}$ from 1 to 1.2, we find that a higher value of $X_{\omega\Delta}$ results a stiffer EOS, which consequently increases the maximum mass. In both cases, no significant change in the canonical radius ($R_{1.4}$) is observed. However, it is found that the change in $X_{\rho\Delta}$ coupling constant keeping constant the value of $X_{\sigma\Delta}$ affects the canonical radius $R_{1.4}$ in a significant way. For example when we fix $X_{\sigma\Delta}$ to 1.1 and change $X_{\rho\Delta}$ from 0 to 1, the canonical radius $R_{1.4}$ changes approximately 1 km. But when we fix $X_{\sigma\Delta}$ to 1.2 and change $X_{\rho\Delta}$ from 0 to 1 then the canonical radius $R_{1.4}$ changes to approximately 1.7 km. An increase in the value of $X_{\rho\Delta}$ leads to an increase in both the maximum mass and the canonical radius, as shown in Fig.\ref{mr}. It also observed that with the coupling constants set at $X_{\sigma\Delta}$ = 1.1 - 1.2, $X_{\omega\Delta}$ = 1, and $X_{\rho\Delta}$ = 0, the NLD parameter set satisfies all constraints of NICER and GW170817.
\begin{figure}
	\includegraphics[width=1.1\columnwidth, height=10cm]{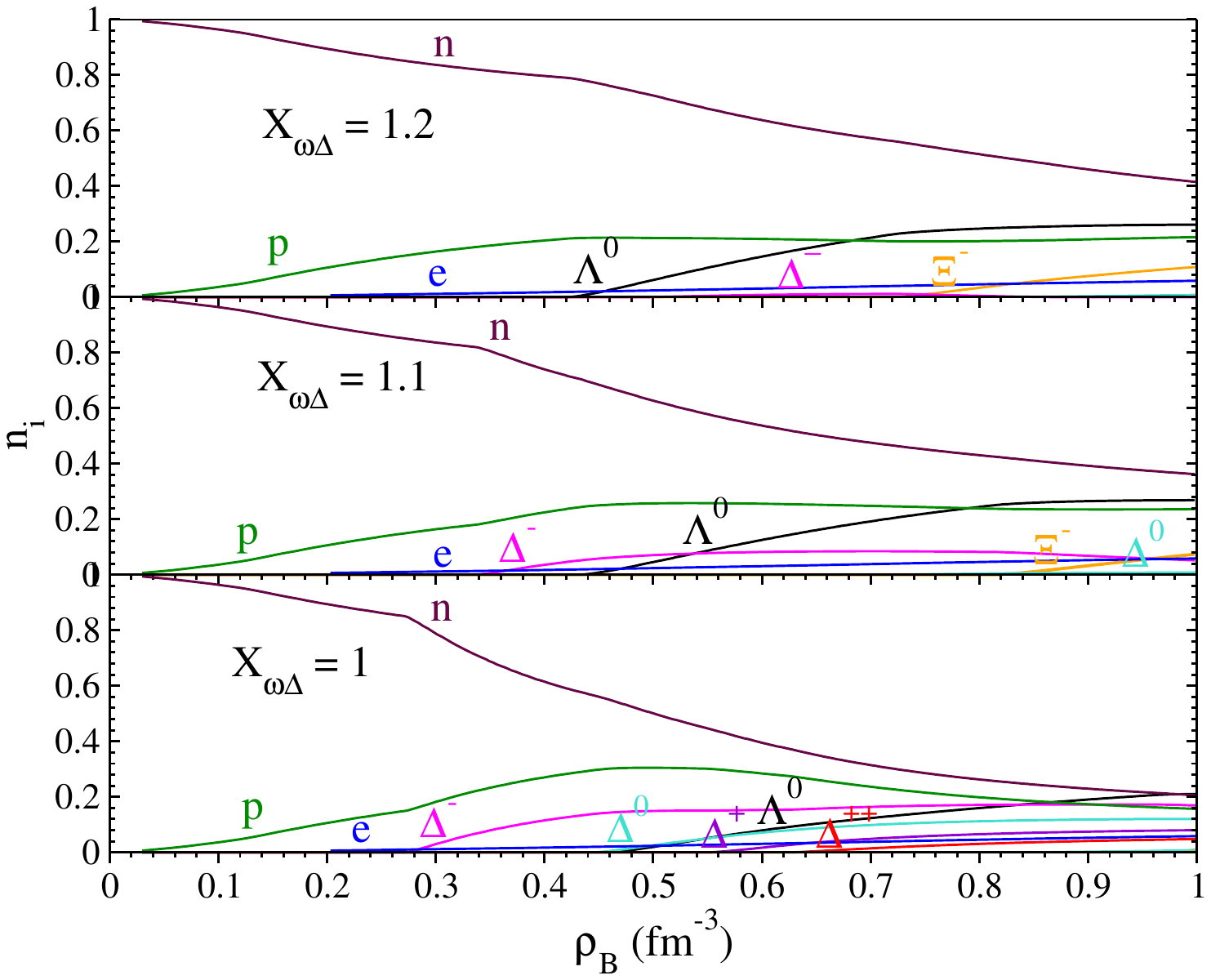}
   \caption{ Baryon formation with different values of X$_{\omega\Delta}$ with X$_{\sigma\Delta}$ = 1.2 and X$_{\rho\Delta}$ = 1  of NLD
parameter set.}
\label{bfw}
\end{figure}
For better clarification, in Fig.\ref{fig4}, we analyze how the canonical radius varies with the coupling constants of the $\Delta$--isobars. Initially, we set the values of $X_{\omega\Delta}$ = $X_{\rho\Delta}$ = 1 and vary $X_{\sigma\Delta}$ from 1 to 1.2, as shown in the left panel of Fig.\ref{fig4}. The results indicate that the canonical radius decreases from 13.690 km to 13.675 km as $X_{\sigma\Delta}$ increases from 1 to 1.2. This reduction is due to the softer equation of state (EOS) for higher $X_{\sigma\Delta}$ values. 
In the middle panel of Fig.\ref{fig4}, we plot the canonical radius ($R_{1.4}$) against $X_{\omega\Delta}$, with $X_{\sigma\Delta}$ = 1.2 and $X_{\rho\Delta} = 1$ held constant. As $X_{\rho\Delta}$ increases, the EOS becomes stiffer (see Fig.\ref{fig3}), resulting in an increase in canonical radius  from 13.675 km to 13.690 km. The right panel of Fig.\ref{fig4} illustrates the variation of canonical radius with $X_{\rho\Delta}$, while keeping $X_{\sigma\Delta}$ = 1.1 and $X_{\omega\Delta}$ = 1. As $X_{\rho\Delta}$ increases from 0 to 2, the EOS also becomes stiffer, leading to an increase in $R_{1.4}$ from 12.68 km to 13.69 km.
From this analysis, it is clear that there is a slight change in $R_{1.4}$ due to change in $X_{\sigma\Delta}$ and $X_{\omega\Delta}$.
\begin{figure}
\begin{subfigure}
  \centering
  \includegraphics[width=1.05\linewidth, height=7cm]{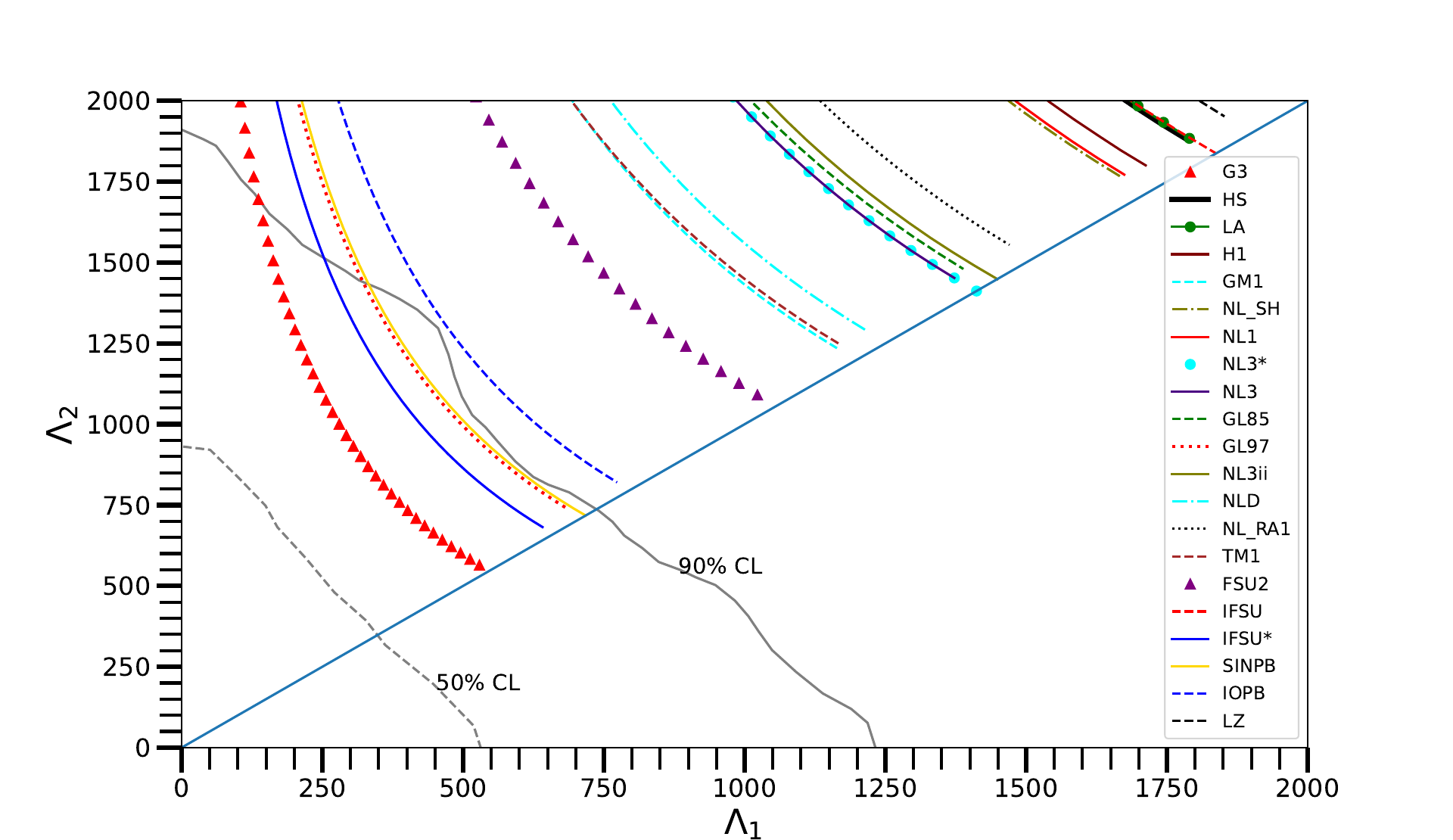}
  \includegraphics[width=1.1\linewidth, height=7.2cm]{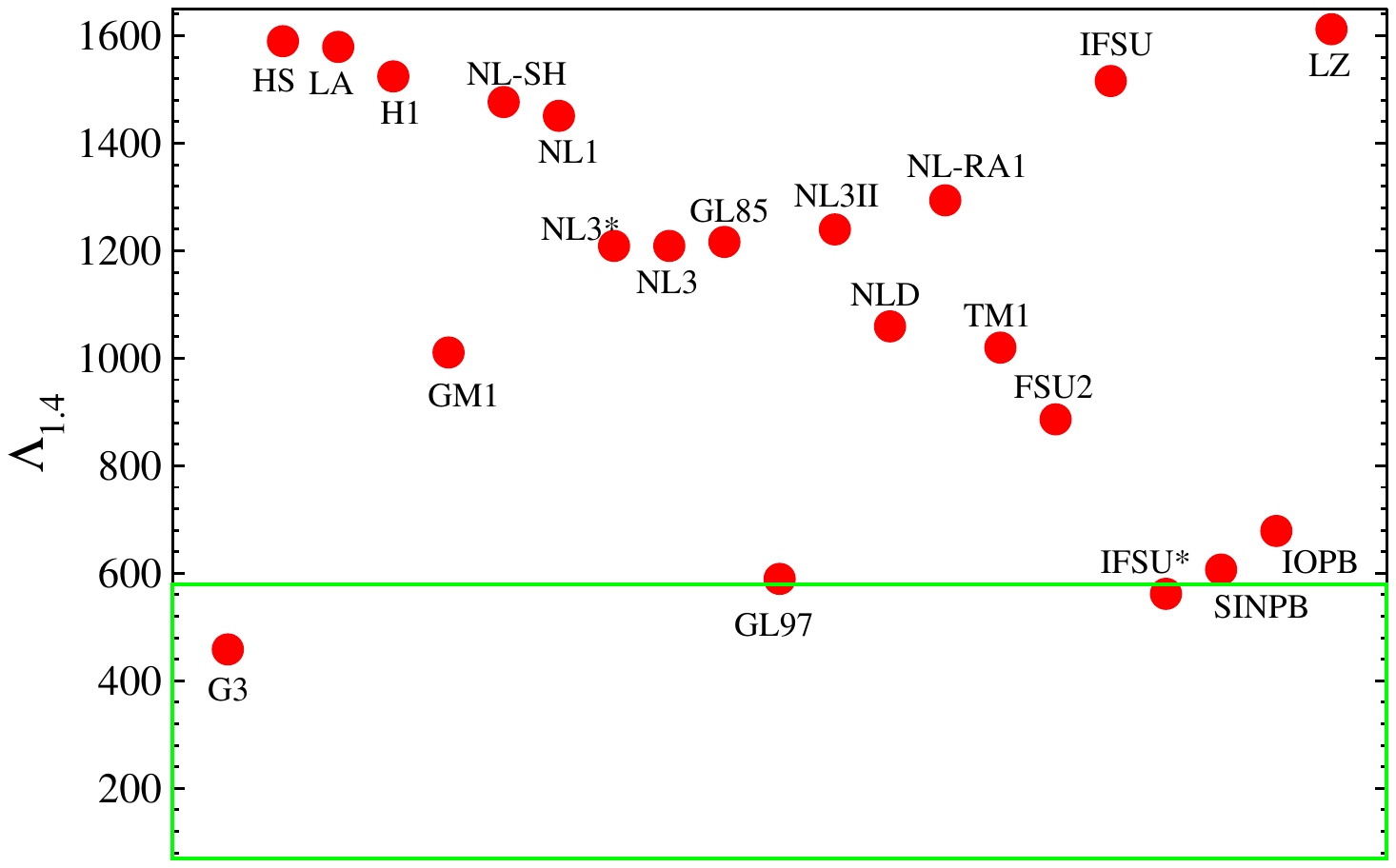}
\end{subfigure}
\caption{Upper figure shows the $\Lambda_1$--$\Lambda_2$ plot of neutron star for various parameter sets of RMF model. 
The solid gray line represents the 90\% credible limit while the dash gray line represents the 50\% credible limit. The diagonal solid line represnt the boundary for $\Lambda_1$ = $\Lambda_2$. The 
lower figure shows the dimensionless canonical tidal deformability $\Lambda_{1.4}$ of neutron star for various 
parameter sets of RMF model. 
The box represent the region for  $\Lambda_{1.4}$ that observed  in the Ref. \cite{abbo18}.}
\label{fig9}
\end{figure}
\subsection{Effect of $\Delta$--isobars on threshold density of hyperons}
In the terrestrial laboratories, hyperons are unstable and decay into nucleons through weak interactions. In contrast, neutron stars have an equilibrium that enables the formation of stable hyperons in their cores at 2--3 times the nuclear saturation density ($\rho_0$). At this density, the chemical potential of nucleons becomes sufficient to convert them into hyperons. The formation of hyperons reduces the Fermi pressure, which softens the EOS of the neutron star. This softer EOS results in a decrease in the maximum mass of the neutron star. The change in the maximum mass is influenced by the nature of interactions between the baryons. If the interactions are attractive, the maximum mass decreases; conversely, if the interactions are repulsive, the maximum mass increases.

In the case of the hyperon star, the $\Lambda^0$ hyperon is a neutral hyperon, and it is lighter in comparison to other hyperons. Also, it  has attractive potential at nuclear saturation density. So, $\Lambda^0$ generally appears at a lower density in comparison to the other hyperons. Though $\Xi$ --hyperons have higher mass in comparison to $\Sigma$ --hyperons, they have attractive potentials at nuclear saturation density. Due to this, $\Xi^-$ appears at slightly higher density after $\Lambda^0$. 
But the $\Sigma^-$ hyperons appear at very high density due to their repulsive potential at nuclear saturation density. When $\Delta$--isobars are included with certain values of $X_{\sigma\Delta}$ and $X_{\omega\Delta}$, $\Delta^-$--isobars appear comparatively lower density than $\Lambda^0$ by replacing a neutron and an electron at top of the Fermi sea. Also, the negatively charged $\Delta^-$--isobar has stronger attractive potential, which balances their mass difference and emerges the $\Delta^-$ at a density before the threshold density of $\Lambda^0$. The $\Delta^0$, $\Delta^+$ and $\Delta^{++}$ isobars generally appear at higher densities just like the positively charged hyperons. A brief description of the population of $\Delta$--isobars at different coupling constants are analyzed.
\begin{figure}
\begin{subfigure}
  \centering
  \includegraphics[width=1.05\linewidth, height=7cm]{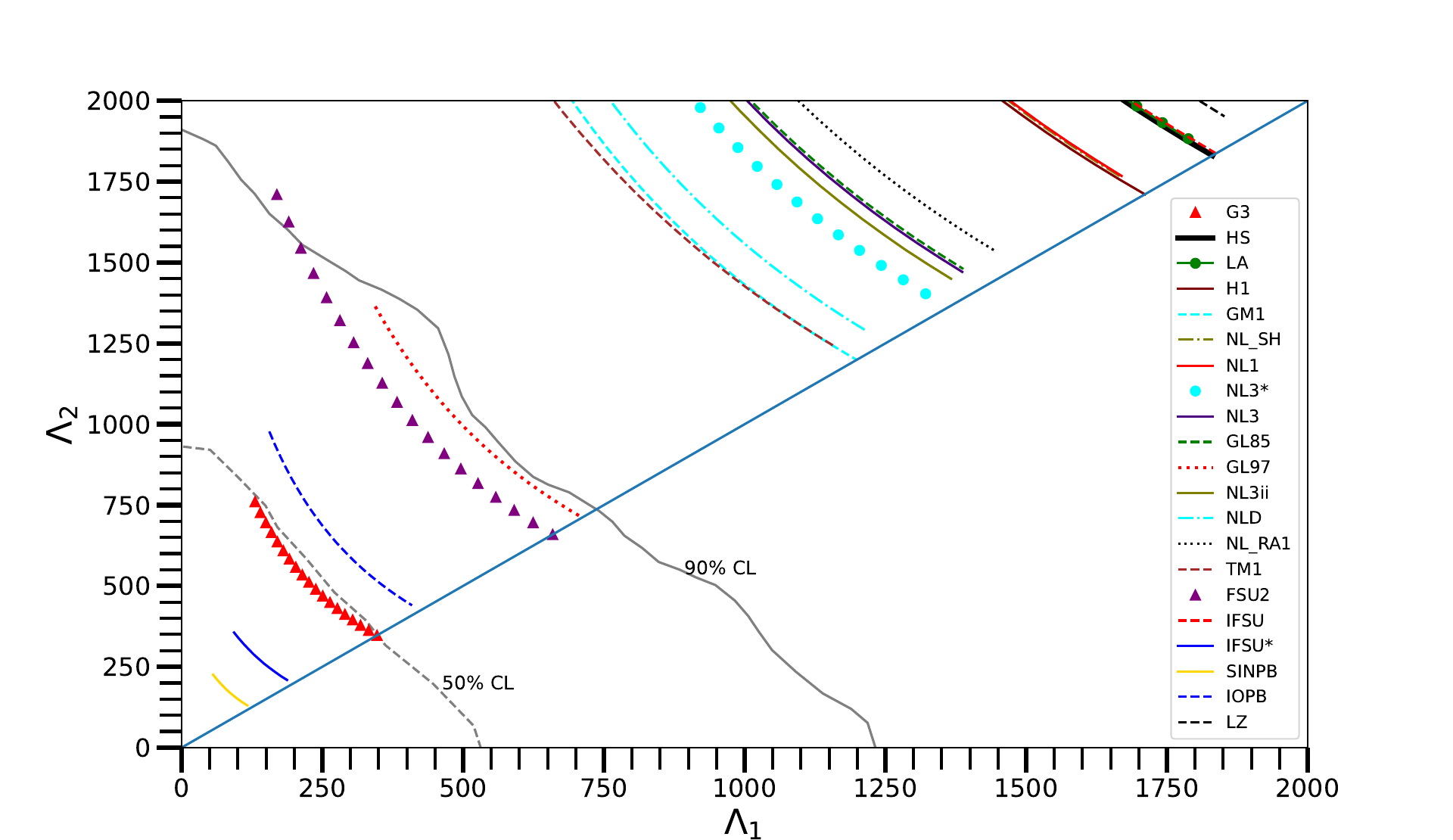}

  \includegraphics[width=1.1\linewidth, height=7.2cm]{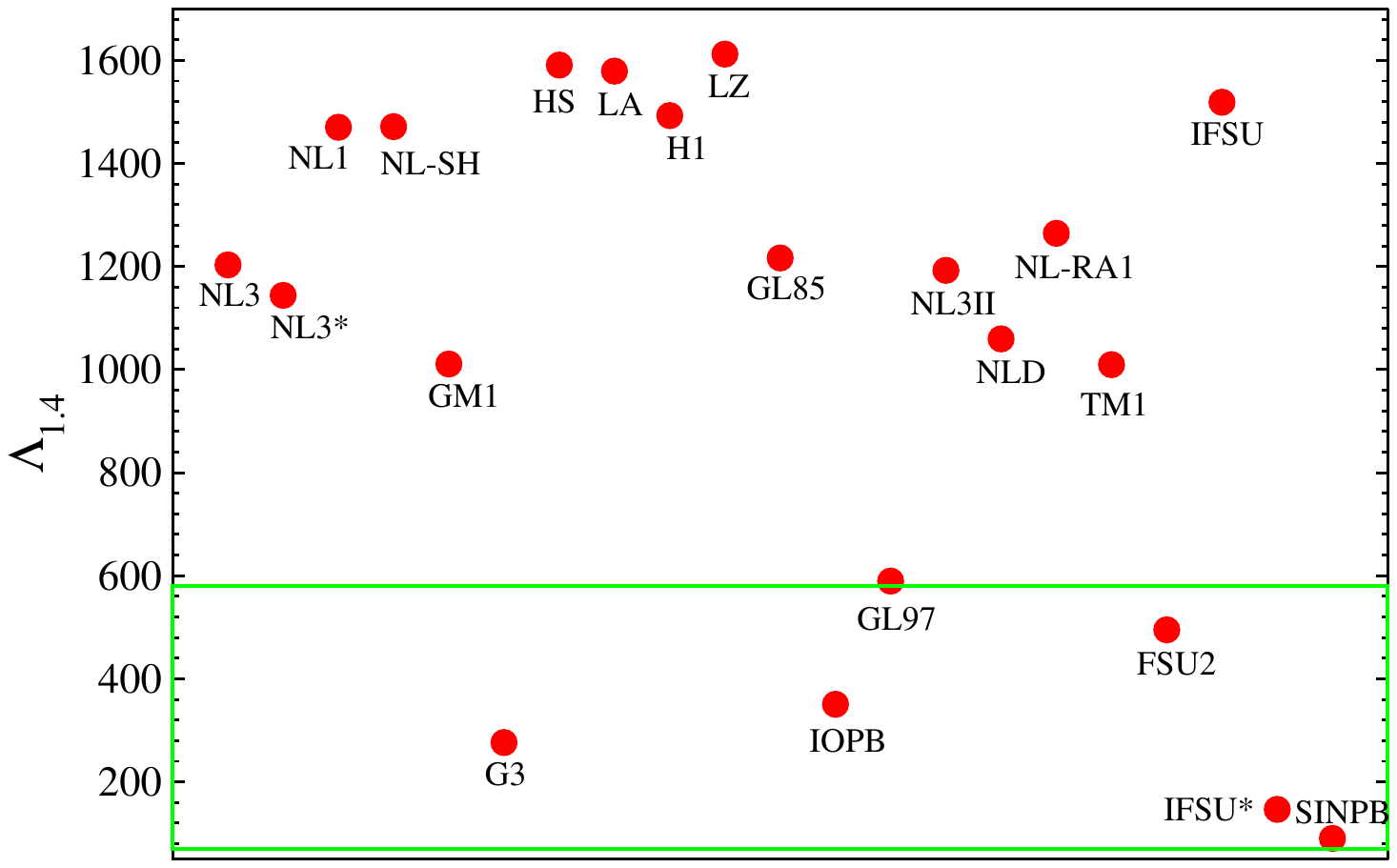}
\end{subfigure}
\caption{Upper figure shows the $\Lambda_1-\Lambda_2$ plot of the neutron star for various parameter sets of the RMF model with $\Delta$--isobars. The solid gray line represents the 90\% credible limit while the dash gray line represents the 50\% credible limit. The diagonal solid line represents the boundary for $\Lambda_1 = \Lambda_2$. The lower figure shows the dimensionless canonical tidal deformability $\Lambda_{1.4}$ of the neutron star with $\Delta$--isobar for various parameter sets of RMF model. The box represent the region for $\Lambda_{1.4}$ that observed in the Ref. \cite{abbo18}.}
\label{fig11}
\end{figure}

We set the coupling constants $X_{\rho\Delta}$ = 1, $X_{\omega\Delta}$ = 1, while varying $X_{\sigma\Delta}$  as 1, 1.1, 1.2 to investigate how it affects the hyperon formation. We observe that by increasing the values of  $X_{\sigma\Delta}$  from 1 to 1.2, the population of $\Delta$--isobar at a lower density increases. Also with the increase of $X_{\sigma\Delta}$, the threshold density of the $\Delta^-$--isobar shifted to lower density. For example, at $X_{\sigma\Delta}$ = 1, the $\Delta^-$--isobar appears just after $\Lambda^0$ ($\rho_B$ = 0.422 $fm^{-3}$) at a density of 0.434 $fm^{-3}$, but for $X_{\sigma\Delta}$ = 1.1 and 1.2, the $\Delta^-$--isobar appears before $\Lambda^0$ (see Fig.\ref{fig5}).  $\Delta^-$--isobar and $\Lambda^0$ appear at densities of 0.339 $fm^{-3}$ and 0.436 $fm^{-3}$,for $X_{\sigma\Delta}$ = 1.1, while for $X_{\sigma\Delta}$ = 1.2, they appear at 0.275  $fm^{-3}$ and 0.459 $fm^{-3}$ respectively (see Fig.\ref{fig5}). The appearance of $\Delta$--isobar at lower density pushes other hyperons to populate at higher density.

We analyze the case where we fix the value of  $X_{\sigma\Delta}$ at 1.1 and $X_{\omega\Delta}$ at 1 and vary $X_{\rho\Delta}$ as 0, 1, 2. 
In this case $\Delta^-$ populates at a higher density for a higher value of $X_{\rho\Delta}$ as shown in Fig.\ref{fig5}. We observe that for $X_{\rho\Delta}$ = 0, the $\Delta^-$ appears much earlier than $\Lambda^0$ ($\rho_B$ = 0.699 $fm^{-3}$) at a density of 0.171 $fm^{-3}$, near the nuclear saturation density ($\rho_0$ = 0.148 $fm^{-3}$). For $X_{\rho\Delta}$ = 1, $\Delta^-$ appears before $\Lambda^0$ at a density of 0.275 $fm^{-3}$ and $\Lambda^0$ appears at 0.459 $fm^{-3}$, while for $X_{\rho\Delta}$ = 2, $\Lambda^0$ appears at a lower density 0.422 $fm^{-3}$ much before $\Delta^-$ which appears at a density of 0.557 $fm^{-3}$. The $\Delta^-$ also appear at higher density than $\Xi^-$, $\Delta^0$, $\Delta^+$ and $\Delta^{++}$. From above analysis it seems that the threshold density of $\Delta^-$--isobar monotonically increases with the value of $X_{\rho\Delta}$. But $X_{\rho\Delta}$ is directly related to the symmetry energy J. So, we can assume that lower value of J facilitated the formation of $\Delta^-$--isobar at lower density. The detailed calculation about how the threshold density of $\Delta^-$--isobar related with symmetry energy is given in the ref. \cite{Cai15}.

Once again, we investigate how different values of $X_{\omega\Delta}$ affect the baryon population in neutron star. We set $X_{\sigma\Delta}$ = 1.2, $X_{\rho\Delta}$ = 1 and vary $X_{\omega\Delta}$ = 1-1.2. We find, by increasing $X_{\omega\Delta}$ value the population of $\Delta^-$ at lower densities gradually decreases as shown in Fig.\ref{bfw}. For $X_{\sigma\Delta}$ < $X_{\omega\Delta}$, $\Delta^-$ appears at a density before $\Lambda^0$ as the attractive potential of the $\sigma$--field dominates over the repulsive potential of $\omega$--field. But, incase of $X_{\sigma\Delta} = X_{\omega\Delta}$ = 1.2, $\omega$--field has large dominance as compared to $\sigma$--field which leads to repulsive interaction of $\sigma$--field at lower density, for which the appear at higher density after $\Lambda^0$. 
\begin{figure}
\vspace{0.5cm}
\includegraphics[width=1.1\columnwidth, height=8cm]{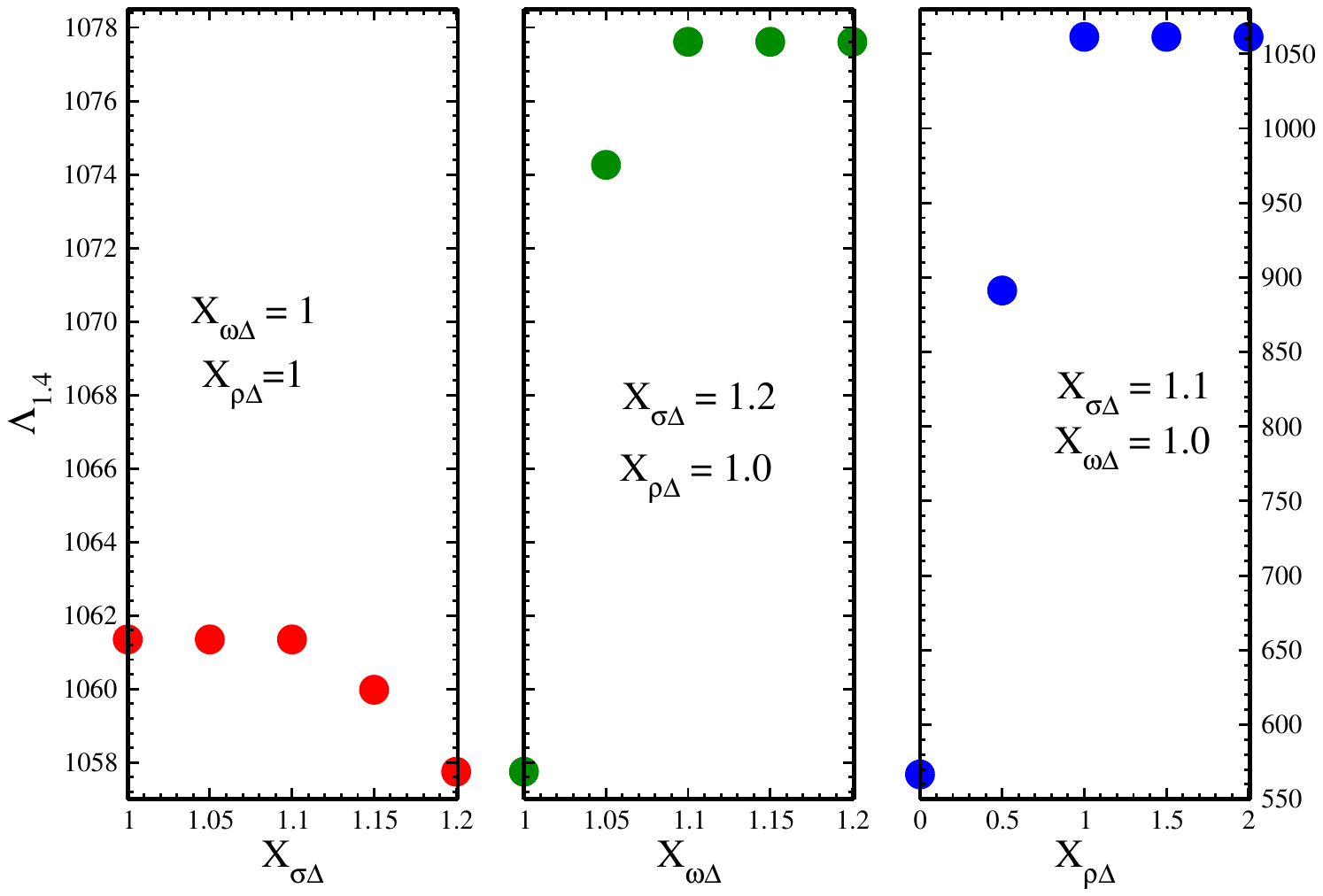}
\caption{Tidal deformability $\Lambda_{1.4}$ of canonical mass of NLD parameter set for various different value of coupling constants X$_{\sigma\Delta}$, X$_{\rho\Delta}$ and X$_{\omega\Delta}$.}
\label{fig13}
\end{figure}

\subsection{Effect on tidal deformability}
In this section, we examine how the inclusion of $\Delta$--isobars influences the dimensionless canonical tidal deformability $\Lambda_{1.4}$ of the neutron star.
We calculate the $\Lambda_{1.4}$ with different RMF parameter sets and observe that the 
parameter sets G3, GL97, SINPB and IFSU* lie within the 90\% credible region, as shown in Fig.\ref{fig9}. These parameter sets have dimensionless canonical tidal deformability $\Lambda_{1.4}$ that lies within the range of $\Lambda_{1.4}$ = 70--580, as reported in the ref. \cite{abbo18}.
From Fig.\ref{fig11}, it is clear that the presence of the $\Delta$--isobars leads to a decrease in the value of the canonical tidal deformability. This observation is further supported by the fact that additional parameter sets, such as FSU2 and IOPB, also fall within 90\% credible limit and the aforementioned canonical tidal deformability range $\Lambda_{1.4}$ = 70--580, when the coupling constants are set to 
$X_{\sigma\Delta} = 1.1$, $X_{\omega\Delta} = 1$, and $X_{\rho\Delta} = 1$, as shown in Fig.\ref{fig11}. The presence of $\Delta$--isobars makes the EOS of the parameter sets SINPB, G3 and IFSU* softer and they lie below 50\% credible region. After the inclusion of $\Delta$--isobars, the canonical tidal deformability $\Lambda_{1.4}$ of FSU2 and IOPB parameter sets reduce and fit within the range of $\Lambda_{1.4}$ reported in the ref. \cite{abbo18}, shown in Fig. \ref{fig11}.

 Further, we calculate $\Lambda_{1.4}$ by changing $X_{\sigma\Delta}$ = 1, 1.1, and 1.2, while keeping $X_{\omega\Delta}$ = $X_{\rho\Delta}$ = 1. It is found that when X$_{\sigma\Delta}$ increases from 1 to 1.2, the value of $\Lambda_{1.4}$ decreases from 1061.35 to 1057.75 as shown in Fig.\ref{fig13}. For higher values of X$_{\sigma\Delta}$, $\Delta^-$--isobar appears at lower density (2--3 times of $\rho_0$), due to this the value of $\Lambda_{1.4}$ decrease.
Furthermore, we change $X_{\omega\Delta}$ from 1 to 1.2 keeping fix $X_{\sigma\Delta}$ = 1.2 and X$_{\rho\Delta}$ = 1. This results in an increase in $\Lambda_{1.4}$ from 1057.75 to 1077.61. The increasing value of $X_{\omega\Delta}$ leads to $\Delta$ appearance at higher densities, causing the stiffening of the EOS, for which the value of $\Lambda_{1.4}$ increases.
To illustrate the effect of interaction strength of $\Delta$--isobars with $\rho$--meson on canonical tidal deformability, we vary the $X_{\rho\Delta}$ from 0 to 2  keeping $X_{\sigma\Delta}$ and $X_{\omega\Delta}$ value fixed at 1.1 and 1, respectively. The result shows that, with increase of value of $X_{\rho\Delta}$, the $\Lambda_{1.4}$ increases from 566.641 to 1061.35 as the $\Delta^-$--isobar populate at higher densities for larger value of X$_{\rho\Delta}$ as shown in Fig.\ref{fig13}.

\section{Conclusions}

In this manuscript, we examine the influence of the $\Delta$--isobar on the bulk properties of the neutron star, like maximum mass, canonical radius $R_{1.4}$ and tidal deformability with the RMF model. There are large number of the RMF parameter sets, but it is not possible to study the detail calculation with all the parameter set. Therefore, we take NLD parameter set as a representative for the detail calculation as NLD parameter satisfy all the constraints reported by NICER and GW170817. From the calculation, it is clear that the production of $\Delta$--isobar at lower density affect the threshold density of the other hyperon. Mostly it affect the production of $\Lambda^0$ hyperon and pushes the threshold density of $\Lambda^0$ to a higher density.

We varies the interaction strength of $\Delta$--isobar with meson keeping in mind the necessary condition that -150 MeV$\leq U_{\Delta} \leq$ -50 MeV and $0\leq X_{\sigma\Delta} - X_{\rho_\Delta}\leq 0.2$. It is concluded from the analysis that the threshold density of the $\Delta$--isobars is strongly affected by the value of $X_{\sigma\Delta}$, $X_{\omega\Delta}$ and $X_{\rho\Delta}$. Fixing these coupling constants in a suitable range we can modify the threshold density of the $\Delta$--isobars within 2--3 times the saturation density. This give an argumentative justification to include $\Delta$--isobars in the calculation of various bulk properties of the hyperon star. For certain values of $X_{\sigma\Delta}$, $X_{\omega\Delta}$ and $X_{\rho\Delta}$, the threshold density of the $\Delta^-$--isobar lies below the threshold density of the $\Lambda^0$.

Our investigation reveals that the $\Delta$--isobars significantly modulate the equation of state (EOS) of the neutron star. The EOS becomes softer for higher value of $X_{\sigma\Delta}$ consequence to decrease in maximum mass and $R_{1.4}$. The increasing value of $X_{\omega\Delta}$ and $X_{\rho\Delta}$ leads to a stiffer EOS, for which maximum mass and $R_{1.4}$ also increase. The most noticeable change in $R_{1.4}$ is observed by changing $X_{\rho\Delta}$ from 0 to 1. The $\Delta$--coupling constants also affect the tidal deformability of the NS. The higher value of $X_{\sigma\Delta}$ decreases the $\Lambda_{1.4}$ while $\Lambda_{1.4}$  increases with increase in $X_{\omega\Delta}$ and $X_{\rho\Delta}$. For $X_{\rho\Delta}$ = 0, the $\Lambda_{1.4}$ vale lies within the range of GW170817 constraint. 

\section{Acknowledgments}
S. K. Biswal acknowledges the support from the SERB, DST,  Government of India, with grant no. CRG/2022/005378.
\section{Reference}
\bibliographystyle{iopart-num}
\bibliography{rashmi}
\end{document}